\documentstyle[12pt,epsf]{article}
\def\appendix#1{
  \addtocounter{section}{1}
 \setcounter{equation}{0}
  \renewcommand{\thesection}{\Alph{section}}
 \section*{Appendix \thesection\protect\indent \parbox[t]{11.715cm} {#1}}
  \addcontentsline{toc}{section}{Appendix \thesection\ \ \ #1}
  }

\def\bea{\begin{eqnarray}}
\def\eea{\end{eqnarray}}
\def\be{\begin{equation}}
\def\ee{\end{equation}}

\def\d{\partial}
\def\D{\delta}

\newcommand{\rf}[1]{(\ref{#1})}
\newcommand{\non}{\nonumber \\*}
\def\t{\theta}\def\ep{\epsilon}\def\a{\alpha}

\begin{document}
\begin{center}
\vspace{32pt}
{\large \bf Phase transition in Wilson loop correlator from AdS/CFT
correspondence}

\end{center}

\vspace{18pt}

\begin{center}
{P. Olesen${}^{a}$ and K. Zarembo${}^{bc}$}

\end{center}

\vspace{18pt}

\begin{tabbing}

{}~~~~~~~~~~~~~~~~~\= blah  \kill
\> ${}^a$ The Niels Bohr Institute, Blegdamsvej 17\\
\> ~~DK-2100 Copenhagen \O .  E-mail: {\tt polesen@nbi.dk} \\
\\
\> ${}^b$ Department of Physics and Astronomy \\
\>~~and Pacific Institute for the
Mathematical Sciences \\
\> ~~University of British Columbia, 6224 Agricultural Road \\
\>~~Vancouver, B. C. Canada V6T 1Z1.  E-mail: {\tt zarembo@physics.ubc.ca}\\
\\
\> ${}^c$ Institute for Theoretical and Experimental Physics, B. 
Cheremushkinskaya 25 \\
\> ~~117269 Moscow, Russia

\end{tabbing}

\vspace{18pt}

\begin{center}

{\bf Abstract}

\end{center}

\bigskip
 
A previous calculation of the phase transition in the Wilson loop correlator
in the zero temperature AdS/CFT correspondence is extended to the case where 
the loops are concentric circles of
unequal radii. This phase transition occurs due to the
instability of the classical string stretched between the loops. We
compute the string action and its expansion in the distance $h$ 
between the loops for small $h$. We
also find that the connected minimal surface is subleading or does not
even exist when $h=0$ and the radii are considerably different. This
feature has no analogue in flat space.
\vfill
\newpage

The AdS/CFT correspondence \cite{ads} allows one to calculate the Wilson 
loop correlator
from  the classical string
action, with the string propagating in the bulk of
$AdS_5\times S_5$ and the ends attached to the Wilson loops lying on the 
boundary \cite{12}. It was pointed out by Gross and Ooguri \cite{gross}
that in the zero temperature case there exists a kind of phase transition, 
corresponding to a competition
between two saddle points. These correspond to a minimal surface
which is either an annulus or two disconnected pieces living at the
individual loops. In \cite{z} the transition between the two phases was 
discussed in detail by solving the equations of motion for the case
when the two loops are circles of equal radii. The result was that at large 
distances $h$ between the loops the disconnected surfaces become
energetically preferred.

In the present paper we consider the case where the two loops have in
general different radii $R_1$ and $R_2$. We find that the equations 
for the minimal surface 
can be solved in terms of the elliptic integrals. 
We construct an expansion in $h$ for small $h$\footnote{This expansion
can be used to get an insight in the structure of electric flux at
short distances \cite{GO}. The structure of electric flux
at large distances was discussed in \cite{ESZ}.}, 
and discuss the critical values of the parameters. In particular we find that
the connected minimal surface is subleading or does not exist if $h=0$ and 
$R_1$ and $R_2$ are considerably different. This feature has no analogue in 
flat space, where the annulus always has smaller area than the two disks if
the Wilson loops lie in the same plane.
 
We should mention that non-analytical behavior in semiclassical
amplitudes related to the transition from one saddle point in the
path integral to another was encountered in many other problems, such
as a false vacuum decay in quantum mechanics,
where both first and second order transitions
are possible depending
on parameters of the problem \cite{chu92}, or 
sphaleron transitions in quantum field theories
with multiple vacua  \cite{hmt}. 
A minimal surface can be regarded as a world sheet instanton and,
in this respect, the phase transition from connected to disconnected
surfaces is a generic example of the  phenomenon common to many
instanton amplitudes.

\section{The equations of motion}

We will calculate the connected correlation function of two Wilson loops
in ${\cal N}=4$ supersymmetric Yang-Mills theory:
\be
\left\langle W(C_1)W(C_2)\right\rangle_{\rm conn} 
=\left\langle W(C_1)W(C_2)\right\rangle-
\left\langle W(C_1)\right\rangle\left\langle W(C_2)\right\rangle,
\ee
where $C_1$ and $C_2$ are concentric
circles of radii $R_1$, $R_2$ that lie in the parallel planes separated
by distance $h$. At strong coupling, the problem reduces to 
computation of the minimal area of a surface in $AdS_5$ whose boundaries are
$C_1$ and $C_2$:
\be
\left\langle W(C_1)W(C_2)\right\rangle_{\rm conn} 
=\exp\left(-\frac{1}{2\pi}\,\sqrt{g^2_{YM}N}\,S\right),
\ee
\be\label{mal2}
S=\int d^2\sigma,\sqrt{\det_{ab}g_{\mu\nu}\d_a x^\mu\d_b x^\nu},
\ee
\be\label{mal2.5}
ds^2=\frac{1}{z^2}(dz^2+dx_\mu^2)
\ee
\be\label{mal3}
\frac{\D S}{\D x^\mu}=0.
\ee
In the given geometry,
the minimal surface is axially symmetric and we can choose the coordinate
along the symmetry axis and the polar angle as the two parameters
on the world sheet. Minimization of the area then reduces to a 
one-dimensional problem, which can be treated along the lines of 
ref.~\cite{z}.

The area of an axially symmetric surface
in the metric of $AdS_5$
is
\be
S=2\pi\int dx\,\frac{r}{z^2}\,\sqrt{1+(r')^2+(z')^2},
\ee
The equations following from this action admit an integral of motion
due to translational invariance in the $x$
direction:
\be\label{imot}
\frac{r}{z^2}\,\frac{1}{\sqrt{1+(r')^2+(z')^2}}=k.
\ee
This integral allows one to get rid of the square root
factors in the equations for $r$ and $z$ which acquire the following 
simple form:
\bea
r''-\frac{r}{k^2z^4}&=&0, \label{one}
\\*
z''+\frac{2r^2}{k^2z^5}&=&0. \label{two}
\eea

We can find the second integral
by first
rewriting \rf{imot} as
\be
(z')^2+(r')^2+1-\frac{r^2}{k^2z^4}=0, \label{three}
\ee
and then adding to it eq.~\rf{one} multiplied by $r$ and eq.~\rf{two}
multiplied by $z$:
\be
(r^2+z^2)''+2=0,
\ee
which yields, upon double integration,
\be
r^2+z^2+(x+c)^2=a^2.
\ee
Here, $a$ and $c$ are the integration constants that are
determined by the boundary conditions:
\be
z(0)=0=z(h)
\ee
\be
r(0)=R_2,~~~r(h)=R_1.
\ee
The latter are satisfied by
\be
c=\frac{R_2^2-R_1^2}{2h}-\frac{h}{2},
\ee
\be
a^2=c^2+R_2^2.
\ee

The trigonometric parameterization:
 \bea
r&=&\sqrt{a^2-x^2}\cos\t, 
\non
z&=&\sqrt{a^2-x^2}\sin\t, \label{trig} 
\eea
allows to separate variables in equation\rf{three} and, after
some calculations, it takes the form:
\be\label{dif}
\t '=\pm\frac{a}{a^2-(x+c)^2}\sqrt{\frac{\cos^2\t}{k^2a^2\sin^4\t}-1}.
\ee
The explicit solution for the minimal surface is  obtained by
integrating this equation with the
boundary conditions:
\be
\t(0)=0=\t(h).
\ee
Besides, the solution must satisfy an inequality
$0\leq\t\leq\pi/2$, which follows from positivity of the AdS
radial coordinate $z$. It is easy to see that
$\t(x)$ grows at small $x$. Then it reaches a maximum at some $x=x_0$, so the
positive root in \rf{dif} should be chosen for $0<x<x_0$.
For $x>x_0$, the solution
monotonously decreases. Hence, the negative root
must be chosen at $x_0<x<h $. 
By continuity, $\t '(x_0)=0$. Equation~\rf{dif} then fixes $\t(x_0)$:
\be
\t(x_0)\equiv\t_0=\arccos\left(\frac{\sqrt{4k^2a^2+1}-1}{2ka}\right).
\ee
Integration of eq.~\rf{dif} from
$0$ to $x_0$ and from $x_0$ to $h$ gives two equations 
that determine $x_0$ and
$k$:
\bea
\frac12\,\ln\frac{a+x_0+c}{a-x_0-c}-\frac12\,\ln\frac{a+c}{a-c}&=&
ka\int_0^{\t_0}\frac{d\phi\,\sin^2\phi}{\sqrt{\cos^2\phi-k^2a^2\sin^4\phi}}
\non
\frac12\,\ln\frac{a+h+c}{a-h-c}-\frac12\,\ln\frac{a+x_0+c}{a-x_0-c}&=&
ka\int_0^{\t_0}\frac{d\phi\,\sin^2\phi}{\sqrt{\cos^2\phi-k^2a^2\sin^4\phi}}
\eea
Adding these equations and defining
\bea
F(ka)&=&
ka\int_0^{\t_0}\frac{d\phi\,\sin^2\phi}{\sqrt{\cos^2\phi-k^2a^2\sin^4\phi}}
=\frac{ka}{2}\,\int_0^{y_0}
\frac{dy\,y^{1/2}}{\sqrt{(1-y)(1-y-(ka)^2 y^2)}}\non
&=&\frac{ka}{2}\int_0^1 du\frac{\sqrt{\sqrt{1+4(ka)^2(1-u)}-1}}
{\sqrt{u\left(2(ka)^2+1-\sqrt{1+4(ka)^2(1-u)}\right)}~\sqrt{1+4(ka)^2(1-u)}},
\label{kkk}
\eea
we get:
\bea\label{ka}
F(ka)&=&\frac14\,\ln\frac{a+h+c}{a-h-c}-\frac14\,\ln\frac{a+c}{a-c}
=\frac14\,\ln\frac{\left(a+\frac{h}{2}\right)^2
-\left(\frac{R_2^2-R_1^2}{2h}\right)^2}{\left(a-\frac{h}{2}\right)^2
-\left(\frac{R_2^2-R_1^2}{2h}\right)^2}
\non
&=&
\frac12\,\ln\frac{R_1^2+R_2^2+h^2+\sqrt{(R_2^2-R_1^2)^2+h^4+2h^2(R_1^2+R_2^2)}}
{2R_1R_2}\,.\label{fka}
\eea

In the last two steps in (\ref{kkk}) we have made the substitutions $y=
\sin^2\phi$ and $u=1-y-(ka)^2y^2$, respectively. The last form of $F(ka)$ has
the advantage that its derivatives with respect to $ka$ are explicitly finite.

There is an ambiguity in solving the last equation 
for $ka$, because generically there are two roots.
It could imply that there are two minimal surfaces with a given boundary, 
but in fact only one of these surfaces is a true minimum of the string action,
while the other is 
a saddle point. It appears that the true solution has larger $ka$.

In the case of equal radii there is an upper limit on $h$, and the same is 
true in the present case. To see this, let us take $R_1/h=\alpha R_2/h\equiv 
\alpha r$. Then we obtain from
(\ref{ka})
\be
r^2=\frac{1}{2\alpha (1+2 \sinh^2F(ka))-\alpha^2-1}.
\label{mmm}
\ee
{}From the condition $r^2>0$ we obtain
\be
1+2\sinh^2F-2\sinh F\cosh F<\alpha<1+2\sinh^2 F+2\sinh F\cosh F.
\ee
The lowest value of $F$ is zero, obtained for $ka\rightarrow\infty$. In this
case only $\alpha=1$ is possible. From (\ref{mmm}) we then obtain $r\rightarrow
\infty$, and since $r=R_2/h$ this corresponds to $h\rightarrow 0$. In
 general, for some value for $F$ we get from (\ref{mmm})
\be
h=R_2\sqrt{2\alpha (1+2\sinh^2F)-\alpha^2-1}.
\label{h}
\ee
The maximum of $h$ therefore occurs for   
\be
\alpha=R_1/R_2=1+2\sinh^2F,
\ee
and hence
\be 
h_{\rm max}=R_2\sinh (2F(ka)).
\ee
Since $F$ is bounded by $\sinh F\approx 0.52$ for $ka\approx 0.58$, we obtain
the result that $h$ cannot exceed $1.172~R_2$. If $R_1=R_2~(\alpha=1)$ we get
from (\ref{h}) $h=2R_2 \sinh F(ka)$, in accordance with a result
derived in \cite{z}.

\section{Computation of the area}

The area of the minimal surface can be computed with the help of the
equations of motion:
\bea
S&=&\int_0^h dx\,\frac{r}{z^2}\,\sqrt{1+(r')^2+(z')^2}
=\frac{2\pi}{k}\int_0^h dx\,\frac{r^2}{z^4}
=\frac{2\pi}{k}\int_0^h \frac{dx}{a^2-(x+c)^2}\,\frac{\cos^2\t}{\sin^4\t}
\non
&=&2\cdot 2\pi\int_0^{\t_0}\frac{d\t\,\cot^2\t}
{\sqrt{\cos^2\t-k^2a^2\sin^4\t}},\label{area1}
\eea
where the factor of two comes from the two branches of $\t(x)$.

The above unregularized
area is ill defined because of the divergency at
the boundary. It needs to be regularized by
 the shift of the surface
into the interior of AdS:
\be
z(0)=\ep=z(h).
\ee
Then, since
\be
\t=\arctan\frac{z}{r},
\ee
the boundary conditions for $\t$ are
\bea
\t(0)&=&\arctan\frac{\ep}{R_2}\approx\frac{\ep}{R_2},
\non
\t(h)&=&\arctan\frac{\ep}{R_1}\approx\frac{\ep}{R_1}.
\eea
After a change of variables
$$\tan\t=\left(\frac{\sqrt{4k^2a^2+1}-1}{2}\right)^{-1/2}\sin\psi,$$
 the regularized area takes the form:
\be
S=\frac{2\pi(R_1+R_2)}{\ep}-4\pi\,\frac{\a}{\sqrt{\a-1}}\int_0^{\pi/2}
\frac{d\psi}{1+\a\sin^2\psi+\sqrt{1+\a\sin^2\psi}}\,,
\ee
where
\be\label{al}
\a=\frac{1+2k^2a^2+\sqrt{1+4k^2a^2}}{2k^2a^2}.
\ee 

The area appears to be a universal function of the parameter $ka$, which
is determined by geometric data according to eq.~\rf{ka}.

As an example of an application of this result, let us consider the
case when $\alpha\rightarrow\infty$, corresponding to $F$ being small.
Then using the above equations we obtain
\be
S\sim -\frac{16\pi^4}{\Gamma(1/4)^4}~\sqrt{\frac{R_1R_2}{(R_1-R_2)^2+h^2}}.
\ee
Here $R_1\approx R_2$ and $h$ is small.

It is possible to gain some insight in the behavior of the action as a 
function of $h$ for small values of $h$. Starting from (\ref{ka}) we get
\be
F(ka)\approx -\frac{1}{2}\ln \frac{R_2}{R_1}+\frac{h^2}{2(R_1^2-R_2^2)}+
O(h^4),
\label{exp}
\ee  
where we assumed $R_1>R_2$ and
\be
h^2\ll 2R_1^2+2R_2^2~~{\rm and}~~h^2\ll \frac{R_1^2-R_2^2}{2(1+R_2^2/R_1^2)}.
\ee
If $h=0$ we determine a value $k_0$ from
\be\label{unpert}
F(k_0a)=-\frac{1}{2}\ln \frac{R_2}{R_1}.
\ee
Expanding $F$ around this point, and assuming that $k_0a$ does not
correspond to the maximum of $F$, we obtain from (\ref{exp})
\be
(k-k_0)a\approx \frac{h^2}{2(R_1^2-R_2^2)F'(k_0a)}+O(h^4).
\ee
Next we expand the action around the point $k_0a$ and to lowest order we then
obtain (ignoring the infinite part of $S$)
\be
S\approx G(k_0a)+\frac{G'(k_0a)}{2(R_1^2-R_2^2)F'(k_0a)}~h^2+O(h^4),
\label{ff}
\ee
where
\be
G(k_0a)=-4\pi\,\frac{\a}{\sqrt{\a-1}}\int_0^{\pi/2}
\frac{d\psi}{1+\a\sin^2\psi+\sqrt{1+\a\sin^2\psi}}.
\ee
Numerical evaluation shows that the two derivatives in (\ref{ff}) are
both negative, and hence the first order $h^2$ correction is positive.

As a numerical example we can take $k_0a=2.6$, corresponding to
$F(k_0a)\approx 0.34$. From (\ref{unpert}) we then get $R_1\approx
2 R_2$. Furthermore, $F'(2.6)\approx -0.054$ and $G'(2.6)\approx -1.77$,
leading to
\be
S(h)\approx -14.8+21.6~\frac{h^2}{R_1^2}+...,~{\rm with}~~R_1\approx 2R_2.
\ee
Numerical studies of eq.(\ref{ff}) show that the coefficient of $h^2/R_1^2$
is quite sensitive to the ratio $R_1/R_2$.
  
\section{Critical behavior}

\begin{figure}[ht]
\hspace*{5cm}
\epsfxsize=7cm
\epsfbox{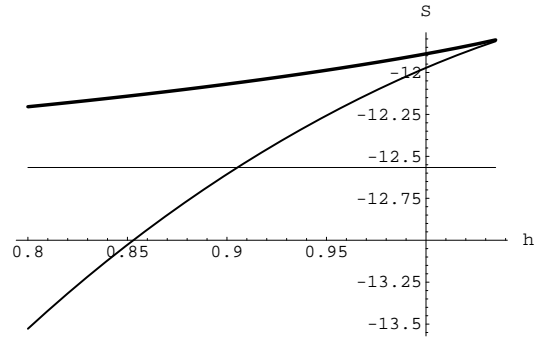}
\caption[x]{The areas of the stable and the unstable
branches (bold lines) of the
connected surface and the area of the disconnected surface 
(thin horizontal line $=-4\pi$) as a functions of
$h$.}
\label{fg1}
\end{figure}

As was mentioned in the discussion following eq.~\rf{fka}, the 
classical string world sheet with the topology of annulus
exists only in a certain range of parameters
$R_1$, $R_2$ and $h$. Outside this range, the minimal surface is disconnected.
Actually, the area of the connected solution starts to exceed the area of
two surfaces that span individual Wilson loops before
the connected solution ceases to exist. In fig.~\ref{fg1}, we plot
the areas of the stable and the unstable branches of the connected solution
as a function of $h$ for $R_1=R_2=1$ vs. the area of the disconnected
surface, which is $-4\pi$ after subtraction of an infinite term, as shown 
in \cite{disc,DGO}.

\begin{figure}[ht]
\hspace*{5cm}
\epsfxsize=7cm
\epsfbox{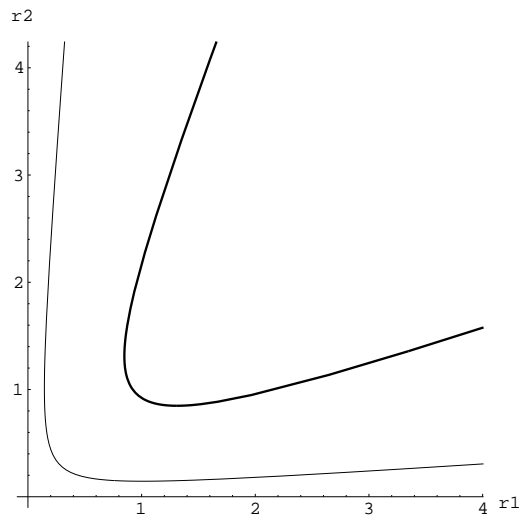}
\caption[x]{The connected minimal surface as function of $r_1=R_1/h$ and
$r_2=R_2/h$ exists to the right of the thin 
line and is the globally stable to the right
of the bold line.}
\label{fg2}
\end{figure}

The shift from one saddle point in the string action to the other leads
to the phase transition in the semiclassical amplitude. In the exact
amplitude,  the transition
is smoothened by corrections that are non-perturbative in 
$\a'=1/\sqrt{g^2_{YM}N}$ \cite{z}, but if $\a'$ is sufficiently small, the
transition is still rather sharp. The phase diagram in the
$r_1$, $r_2$ plane is shown in fig.~\ref{fg2}, where $r_1=R_1/h$
and $r_2=R_2/h$. It is interesting that the phase transition survives
the $h\rightarrow 0$ limit, which corresponds to $r_1,r_2\rightarrow\infty$.
Therefore, the connected minimal surface is subleading or even does not
exist if $h=0$, but $R_1$ and $R_2$ differ considerably. This behavior
has no analogue in the flat
space where the annulus always has smaller area than the two disks if 
the Wilson loops lie in the same plane.

\subsection*{Acknowledgments}

The work of K.Z. was supported by 
NSERC of Canada, by Pacific Institute for the Mathematical Sciences
and in part by RFBR 
grants 98-01-00327 and
00-15-96557 for the promotion of scientific schools.

\end{document}